\documentclass[12pt, preprint]{aastex}
\usepackage{bm}
\begin{document}
\title{Standard GRB, Dirty Fireballs, and the Excluded Middle}

\author{David Eichler \altaffilmark{1}
}\altaffiltext{1}{Physics Department, Ben-Gurion University,
Be'er-Sheva 84105, Israel; eichler@bgu.ac.il}

\begin{abstract}
It is shown that the contribution of faint gamma-ray bursts (GRB) to the total allsky GRB energy flux is   small. The allsky flux of GRB appears to be $5.3 \times 10^{-3}$ erg/$\rm cm^2 y$, with little additional component hidden within weak or otherwise undetectable GRB. This significantly constrains physical  models of GRB and dirty fireballs, suggesting a rather sharp dichotomy between them should the latter exist.
\end{abstract}

\section{Introduction}
  Gamma-ray bursts  (GRB) are understood to depend on baryon purity. With too much baryon contamination,   their energy would end up in  kinetic energy, and eventually afterglow. The rather stringent requirements for baryon purity needed for a successful GRB are consistent with the inference that there is only about 1 GRB per 10,000 supernovae, despite the known association between the two phenomena.

  One might therefore ask whether there is a continuous spectrum of "semi-GRB" that smoothly connects normal supernovae to bright GRB. Dirty fireballs and X-ray flashes have been proposed as examples of this. As X-ray flashes are about as numerous as standard GRB, but have much  smaller  values for $V_{max}$, the maximum value within which they could have been observable at our location, their rate density is probably much higher, and one might ask whether the total luminosity input in such semi-GRB's could dominate the total. This is relevant to any question connected with GRB calorimetry and total luminosity density, e.g. the question of how much energy is available in GRB to produce ultrarelativistic cosmic rays (UHECR).

  There is physical motivation for asking this question as well, as it constrains models of the central engine.  If, for example, the baryon purity is a function of disk cooling, so that baryons from a hot atmosphere do not extend up to the slow-Alfven critical point where they are injected into a magnetocentrifugal outflow (Levinson 2006), then one might suppose that, because this injection rate varies both with position on the disk and time,  there are many GRB in which the photons are cooled by adiabatic expansion before being released beyond the gamma ray photosphere. In fact, popular models of GRB frequently invoke a saturation radius that is well within the photosphere, in which case the photons would cool adiabatically  before escaping. If there is a continuous range of the extent of this adiabatic cooling, then one would expect a continuous range of GRB fluences, and perhaps a sizable contribution to the total allsky fluence from very weak, very soft, or very prolonged (i.e. low luminosity) GRB. Yet another  parameter that seems quite variable in GRB is the opening angle. The larger the opening angle, the dimmer the GRB appears. Very wide GRB might thus escape detection, yet might dominate the total luminosity density of GRB. Similarly, the peripheries of "structured jets" might conceivably  contain most of the GRB energy, in which case different observers would see the peripheries as weak GRB, but with a large rate per unit comoving volume, and the product of the rate density and apparent energy of the weak GRB might conceivably dominate.

  In this letter  we examine to what extent weak, soft, wide angle and/or high redshift GRB contribute to the present cosmic GRB radiation density. In contrast to previous works on this matter, which appear to be model dependent and give a large range of numbers, our approach is nearly model independent - we compute the total flux recorded by different GRB detectors, each with different ranges and sensitivities.  We show that   most of the GRB energy output comes in high fluence GRBs in the spectral range 20-5,000 KeV, independent of detector  and trigger sensitivity, and such bursts could be detected by most major detectors. Moreover, these bright bursts would typically overwrite data that was being transmitted from BATSE, should they have  occurred during the time such data transmission was taking place (considered "dead time" except when overwritten [Paciesas et al., 1999]), and would not have been overlooked by BATSE (see below).

  We conclude from this that, in the context of total energetics, GRBs respect a law of excluded middle: There appears to be very little energy in "semi-GRBs", compromised GRBs,  ragged GRB peripheries, and the like. Theoretical models of GRB should respect this principle of the excluded middle at a quantitative level. In other words, they should not predict that  half or more of the total photon energy is sprayed or otherwise diluted in directions from which the burst would be perceived as a marginal, weak,  or "semi-" GRB. While the detectability of any given weak GRB is a function of detector sensitivity, the measured allsky flux in GRB is insensitive to it.

  We find from the BATSE 4B catalog that the total allsky flux in GRB photons  $F_{\gamma}$ is about $5.3\times 10^{-3}$ erg/cm$^2$yr. {Specifically, we  have summed the fluences of all the GRBs with reported fluences,
  divided  by the number of years (5.3) and, assuming an effective field of view (FOV) of $2\pi$ sr, multiplied by 2 in order to obtain an average allsky  ($4\pi$)                                        flux.
  This flux  corresponds to a cosmic average energy input $F_{\gamma}H/c = 1.24 \times 10^{43}$erg/Mpc$^3$y. The present day, local input is about a factor of 2 lower due to the fact that the star formation rate, and hence the inferred long GRB rate, was considerably higher in the past (e.g. Eichler, Guetta and Pohl 2010), i.e. enough to override the average redshift factor by a factor of 2. These values include short bursts, and for long bursts  would probably be lower by  several percent or so.

  Previously (Eichler, Guetta and Pohl, 2010), we  summed the total fluence in GBM data from August   2009 through February 2010, an 18 month interval, and found an average flux in {\it long } GRB of $4\times 10^{-3}$erg cm$^{-1} \rm y^{-1}$. The GBM catalogue contains 212 long burst over this period, or about 283 GRBs per year per $4\pi$ sr.

  The BATSE 4B catalogue contains 1293 GRBs with fully reported fluences above 20 KeV over 5.3 years. This corresponds to an allsky rate of $\sim 450$ to 500 GRBs per year with measured fluences, about $10^2$ of which are short GRB. This number is constant over subintervals of order 1 year to within $\sim$2 percent. In addition, there are about 70 per year that are interrupted by data gaps,  so no measured fluence is available, and we may infer that a comparable number are not even triggered because they occur during the "dead time" when the satellite trigger is preoccupied  with relaying data from the previous GRB, or in a region of high particle background flux.
  Finally, some GRBs may be go  unreported because the detector is not in its most sensitive trigger mode all of the time. In its most sensitive mode, BATSE is thought to trigger on about 333 GRBs per $2\pi$ sr y, or an all-sky flux of 666 per year  (Paciesas et al 1999). At least $\sim 150$ of these would be short GRB, and we may infer that the number of long GRBs is $\sim 500$ per year.  Band  (2002) infers a somewhat lower allsky rate (550 per year versus 666), which would imply less than 500 long GRBs per year.
     This suggests that a fraction $\eta$ between 18 and 34 percent of GRB in the BATSE field of view went undetected because of dead time effects or because the trigger threshold was not set at optimal levels. We suspect that $\eta$ is closer to 18\% because at most 70/520 per $4\pi$ sr-year of the triggered GRBs were interrupted by overwrites or data gaps.
     On the other hand, the average fluence per reported burst is likely to be stronger than for an interrupted or sub-trigger threshold burst, so, although this mistake is sometimes made, it would be incorrect to multiply the total number of inferred long GRBs by the  average fluence in the proper subset of those with reported fluences, since the latter are brighter as a class than the former.  Previous estimates in the literature for the GRB radiation luminosity density, which are often estimated by multiplying an inferred rate by an inferred average energy, are  for this reason somewhat larger than what we estimate here, and we consider them to be overestimates. The problem is particularly severe given that the total fluence is dominated by bright bursts  average fluence, $F \ge 10^{-5}$ erg cm$^{-2}$, whereas the total {\it number} of bursts is dominated by dimmer  bursts, $F \le 10^{-5}$ erg cm$^{-2}$.

Finally, we consider the event rate recorded by Swift. While the
limited energy range of Swift makes it a bad instrument for
measuring total fluence, it is nearly an order of magnitude more
sensitive than BATSE for merely detecting GRB. It recorded $\sim
93.5$ GRB per year over a field of view of $1.4$ ster. This implies
an allsky event rate of 840 per year, which is 20 - 50  percent more
than the BATSE all-sky event rate. It may be assumed that the extra
bursts are below the BATSE threshold and therefore are extremely
weak.  It may be inferred that an instrument with Swift trigger
sensitivity and BATSE energy range would have recorded
 only a
slightly  higher flux than BATSE itself. It is also
 clear that the
modest increase in burst rate that results from the
large increase
in sensitivity indicates that the total burst rate
converges at
large sensitivity, unless there is a distinctly separate
 class of
bursts at very low flux levels.

In Figure 1 we have plotted the cumulative 4-channel fluence,
 $\sum NF_4(\ge F_4)$ (the fluence in channels 1 though 4,
 20-1000 KeV) for the 1293 GRBs in the BATSE 4B catalogue
 that have a reported fluence,  as a function of 4-channel
 (i.e. total) fluence $F_4$.  Here the cumulative fluence is
 defined to be the sum of the fluences in all bursts brighter
 than F.   We also plot the cumulative  2-channel fluence
 (channels 1 and 2, 20 -100 KeV) for the BATSE 4B catalogue
 as a function of the 2 channel fluence. To assist the eye,
 we plot the cumulative fluence as $6.8\sum NF_2(\ge 6.8F_2)$
 for the BATSE 2-channel  fluences, which expresses the
 simplifying assumption that the total fluence is 6.8 times the 2-channel fluence $F_2$.  The bolometric correction factor of 6.8 is motivated by the fact that it is the average ratio of $F_4/F_2$ for the BATSE catalogue. It can be seen that $6.8 F_2$ is nearly a proxy for $F_4$,  except that the ratio $F_4/F_2$ is somewhat higher than 6.8  for very bright bursts, which are probably seen head-on and therefore have harder than average spectra,   and also higher for short bursts, which have very low fluences, and contribute significantly to the low-fluence population. (That the ratio $F_4/F_2$ is lower in the range $10^{-6}$-$10^{-5}$erg/cm$^2$s than at the extremes is consistent with the fact that moderate fluence bursts have softer spectra than the bright GRB, as implied by the Amati relation.)
Also plotted is the  cumulative fluence vs. fluence as recorded by Swift (15-150 KeV) for 534 GRBs in the Swift catalogue  with (reported fluence) as of Sept 1, 2010  as a function  the Swift-recorded fluence, $\sum NF_s(\ge F_s)$, with bolometric corrections of 3.15 and 1.8 (see figure caption and below).

As Swift has a larger energy range than channels 1+2 of BATSE,  it follows that the bolometric  correction to Swift-measured fluences must be less than 6.8. As the average Swift-measured fluence was $\sim 3.15$ times less than the average BATSE  4-channel fluence, and Swift sampled a fainter population on average, the bolometric correction should probably be less than 3.15. 
Finally, we may argue that because Swift  detects an allsky
equivalent of 840 bursts per year and a total fluence of $1.82
\times 10^{-3}$erg/cm$^2$ in 534 bursts, the implied allsky flux is
$[840/534 ] \times 1.82 \times 10^{-3} =2.9\times
10^{-3}$erg/cm$^2$y. Because Swift, with proper bolometric
correction, should detect a higher flux than the less sensitive
BATSE, it follows that Swift should detect an allsky
equivalent of
at least $ 5.3 \times 10^{-3}$erg/cm$^2$y.  So the proper bolometric
correction must be at least 5.3/2.9=1.8. It is clear from Figure 1
that for any allowed choice of the Swift bolometric correction, the
cumulative fluence plateaus at a fluence of $\sim
10^{-5}$erg/cm$^2$, which is three orders of magnitude from the
minimum. Note that choosing the lower limit for the bolometric
correction would move the plateau to the right slightly, thus
slightly shortening it, but it would also imply that the allsky flux
in Swift bursts is not significantly higher than in BATSE bursts,
which is what the long plateau signifies.

To summarize, the cumulative fluence plateaus near its
maximum value at a fluence that is about a factor of
$\sim 10^3$ higher than the minimum  for both BATSE
and Swift,  suggesting that there is not a significant
amount of flux in weak GRB, at most of order 15\% or so.
A significant part of this can be attributed to short GRB.
While it might have otherwise been argued that there
remains a reservoir of not-yet-detected dim, long GRB,
we would expect, if that were the case, that Swift
would have detected at least part of that reservoir,
and shown more of a rise in cumulative fluence in
the range of low fluence bursts. We conclude that
long GRB and their total luminosity have been fairly
sampled by existing instruments, and that what we
see is what we get.

 Similar remarks can be made for GRBs at high redshift.
 Swift would have been better suited, because of its high
 sensitivity and lower energy range, to see GRB at high
 redshift z.  While the average redshift of GRBs in the
 Swift catalogue is certainly higher than that of the
  BATSE catalogue,  there is no evidence that the
  population of high z GRBs makes any significant
  contribution to the cumulative fluence.  In any
  case, whatever the contribution of GRBs that,
  because of  high redshift, are presently undetectable,
  this contribution would not greatly affect the
  estimate of the present day, local estimates,
  which are usually figured by assuming the GRB
  redshift distribution follows that of star formation.
  This is independent of whether high redshift activity
  is more conspicuous, relative to low redshift activity,
  in young stars or in GRB. Uncertainties in the GRB
  event rate at high z therefore do not affect the
  estimate of the local ($z\sim0$) GRB luminosity density.

We note that this result is consistent with the Amati relation,
which finds that the isotropic equivalent luminosity $E_{iso}$
 goes as the square of the spectral peak energy $E_{peak}$, i.e.
 that the total
photon number decreases in proportion to the spectral peak. If the
number of GRB per logarithmic interval in $E_{peak}$ is roughly
constant, then the number of photons contributed by X-ray flashes is
much less than by the harder GRB.\footnote{In fact, the Amati
relation is actually an inequality [Nakar and Piran 2005] in that
there are GRB with $E_{iso}$ less - but none with $E_{iso}$  much
greater - than the value given by the Amati relation, and the number
of photons may for many bursts decrease even more rapidly with peak
energy.} The Amati relation is thus consistent with spectral
softening due entirely to off-axis viewing (Eichler and Levinson
2004), which reduces the photon flux as well as the photon energy.
It is not consistent with the hypothesis that X-ray flashes are the
same as GRB except for more adiabatic cooling of the photons,
because such adiabatic cooling would leave the total number of
photons unchanged.

We conclude that most of the energy flux, burst rate, and probably
even the photon flux have all been measured to good accuracy, and
that the hidden contribution from weak GRB is small. We find that
the net allsky energy flux in GRB (electromagnetic) radiation is
$\sim 5.3 \times 10^{-3}$ erg$/cm^2 y$, which is is consistent with
Schmidt (2002, private communication to V. Berezinsky) who estimated
a net GRB photon luminosity density  of $6\times 10^{42}\,\rm
erg/Mpc^3 y$.
It is, however, a factor of several below some  estimates
that are
based on nominal average GRB energy as deduced from a
selected
subsample from known redshifts. This may be attributable  to the
fact that  GRBs with known redshifts  have higher fluences on the
average that those of undetermined redshift, and that using the
former as representative of the latter leads to an overestimate..

 We emphasize that  the measured flux applies only to photons. No limit on kinetic energy is placed by such considerations. Thus, if GRBs, including those beamed away from us, occur at a rate of $100/Gpc^3$/y and each GRB has $10^{53}$ erg in kinetic energy, then an energy input rate of $10^{46}$erg/$\rm Mpc^3 y$ is possible; it would simply mean that the kinetic energy input rate exceeds that of photons by  3 orders of magnitude. One might expect some of this kinetic energy to escape as orphan afterglow radiation, however, in some other part of the spectrum, or, if this outflow contains particle-accelerating internal shocks  (Levinson and Eichler, 1993), even UHE neutrinos (Eichler, 1994).

That the GRB population has a reasonably sharp edge in the context
of cumulative fluence does not rule out the possibility that most
GRB,  as counted by external observers,  are dirty fireballs, and/or
that the dirty fireball component of a GRB has a much larger solid
angle than the radiative core of the fireball. However, it
constrains the dirty fireballs to be extremely dirty - qualitatively
different from the radiative core, rather that there being a
continuous transition between the two.  In fact, we have argued in a
previous paper (Mandal and Eichler, 2010) that the thermal component
of GRB 060218 is a dirty fireball, as it is the only known exception
to the Amati relation that lies to its bright, soft side (the lower
right as the Amati relation is usually plotted). The non-thermal
 component of this GRB does in fact respect the Amati relation
 (Campana et al 2006, Amati et al 2007), as it has
 comparable or only slightly more  isotropic equivalent energy
 ($\sim 4.5 \times 10^{49}$erg) than the thermal
 component($\sim 3 \times 10^{49}$erg) while having a higher
 spectral peak ($\sim 4.9$ KeV as opposed to
 $\sim 0.486$ Kev, the Wein peak at a temperature of 0.17 KeV).\footnote{From figure 2 of Campana et al (2006), the
 0.3-10 KeV luminosity is
 $ \sim 7.6 \times 10^{49}$ erg/s, and the thermal component is $\sim 3.1 \times 10^{49}$ erg/s.
 The authors state a luminosity of $\sim 6.2 \times 10^{49}$ erg/s
 above 1 KeV.}
 It has also been noted
 (e.g. Guetta and Eichler 2010)\footnote{In Guetta \&
 Eichler (2010)  there is a typo as the sky coverage
 of Swift is not 0.17 sr but rather 0.17} that the event
rate density of GRB 060218-type events is more than two orders of
magnitude higher than that of standard GRBs. However, the thermal component of GRB 060218 has
a spectral peak at $\sim 0.5$ keV,
  which is more than 3 orders of magnitude below that of
  the brightest GRB, and is at least three orders of
  magnitude  dimmer, in terms of $E_{iso}$, than the
  brightest GRB.
There do not seem to be any GRBs on a line connecting GRB 060218
with the brightest GRB lying in between the two extremes - which
would correspond to less dirty fireballs - despite the fact that
such GRBs would, if they existed, have a much higher $V_{max}$ than
that of GRB 060218, and that they would be  brighter than GRB of
comparable spectral peak that do respect the Amati relation (of
which there are many). This would, under the assumption that the
thermal component of GRB 060218 is a dirty fireball, support the
picture that dirty fireballs are extremely dirty and that there is
no known class of objects that parametrically connects dirty
fireballs to standard GRB. Figure 2 shows that the thermal component
of GRB 060218 is the only burst or flash to lie to the lower right
of the line depicting the Amati relation. It connects to the bright
GRB in the upper right along a line of constant photon density,
which would correspond to adiabatic cooling of a typical GRB photon
output by an excessive amount of baryonic contamination that traps
the photons in the expanding flow, but there are no known GRB in
between the bright GRB and GRB 060218 on this line.

What are we to make of this sharp dichotomy between standard GRB and dirty fireballs?   One obvious interpretation is that long GRB are collimated by their host stars and have sharp edges. Apparently, very little of their energy escapes  into a peripheral spray that would be observed as weak GRB. It would be interesting to do a similar analysis on short GRB to see if a similar effect obtains, given that short GRB do not emerge from giant envelopes because their host stars do not possess them.  Such an analysis seems only marginally possible at present, because short GRB are of much lower fluence  and rarer than long GRB, but it should eventually be feasible.

If dirty fireballs (be they material from the inner accretion disk
of the GRB's central engine or entrained envelope material) ensheath
the radiative cores of GRB fireballs,  then it appears that they do
not parametrically connect smoothly to the radiative core. They may,
rather,  connect smoothly to the supernova ejecta.  This would be
consistent with the picture that standard GRB fireballs are very
uncontaminated by baryons because they emerge along field lines that
connect to an event horizon (Levinson and Eichler, 1993), and that
the dirty fireball emerges along field lines that connect to an
accretion disk. The "either/or" nature of the event  horizon then
creates a large parametric gap, nearly a discontinuity,  between the
baryon content of the radiative core and the dirty fireball.

\begin{figure}
\includegraphics[width=01.0\columnwidth, keepaspectratio]{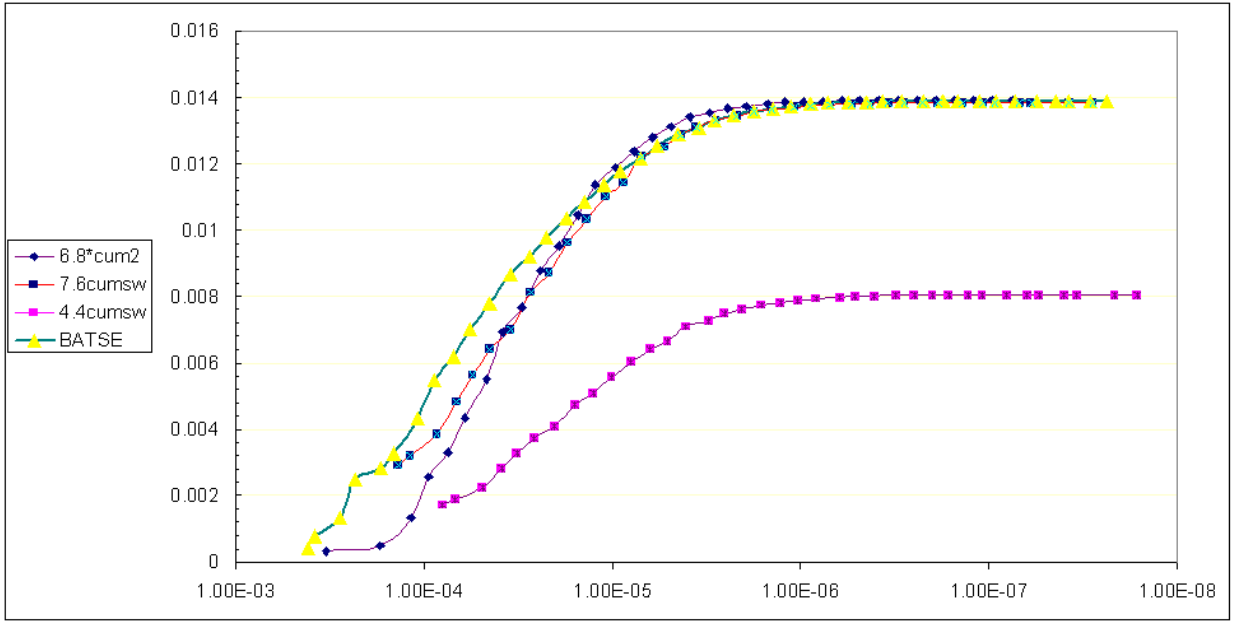}
\figcaption[cumfl.eps]{The cumulative fluence for the
four BATSE energy channels, labeled BATSE, is plotted as
a function of 4-channel fluence. The curve labeled 6.8cum2
plots the cumulative BATSE fluence in channels 1 and 2 as a
function of the 2-channel fluence, but with a bolometric
correction of 6.8 for the 1+2 channel fluence, and it is
clearly very close to the 4-channel fluence, indicating
that this is the proper bolometric correction for the
restricted energy range of channels 1 and 2 (20-100 keV).
 The curve labeled 7.6cumsw is the cumulative fluence as
  a function of fluence, both as measured by Swift, but
  with a bolometric correction of 3.15 applied to the
  Swift-measured  fluence, and a normalization factor
  of 1293/534 to account for the fact that the BATSE
  (Swift) data set contained 1293  (534) GRB. The curve
  labeled 4.4cumsw is the same but with a bolometric
  correction of 1.8 and the same 1293/534 normalization
  factor. The units of fluence are erg/cm$^2$.}{\label{f01}\footnotesize}
\end{figure}

\begin{figure}
\includegraphics[width=1.0\columnwidth, keepaspectratio]{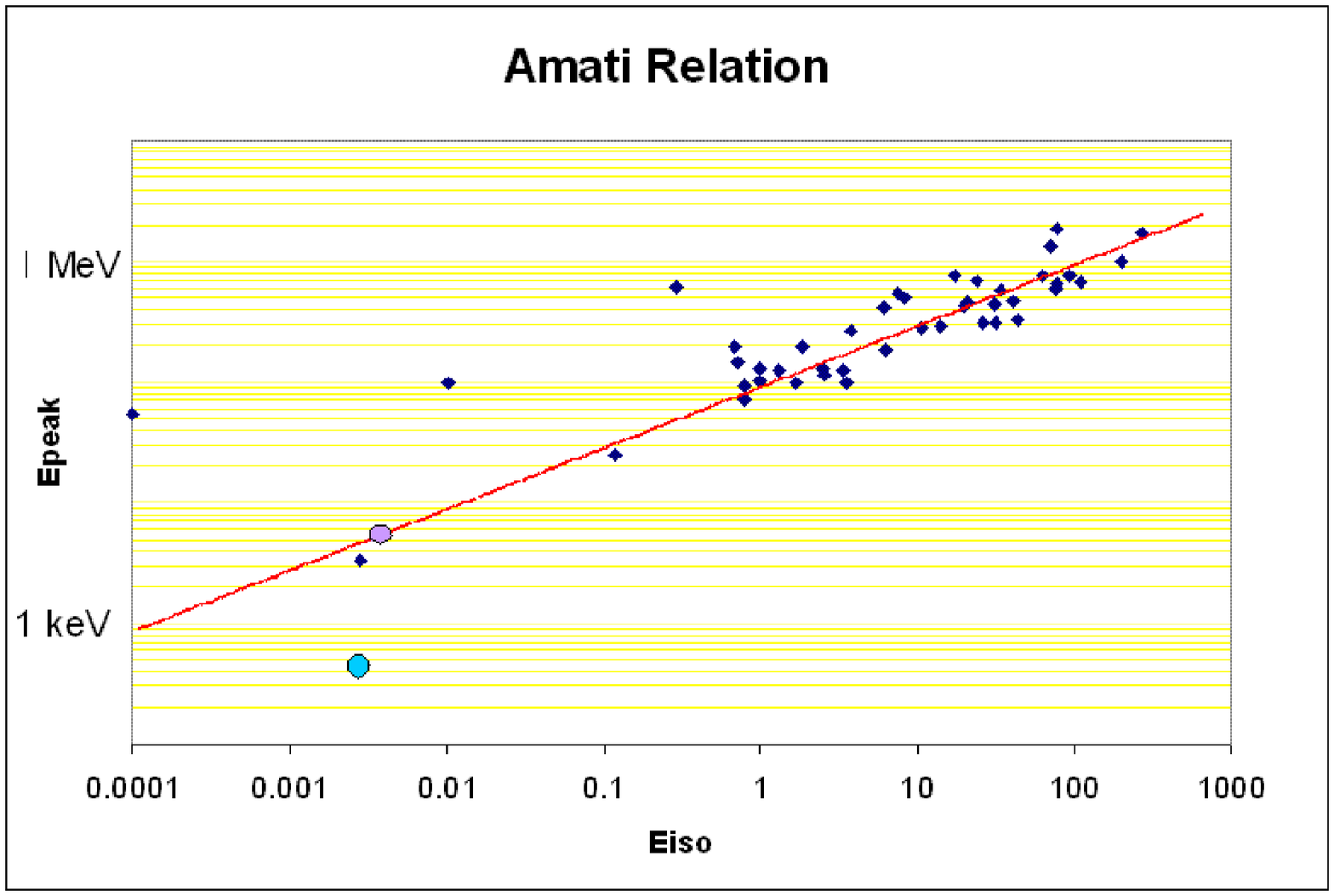}
\figcaption[amati.eps]{The Amati relation is illustrated with data
taken from Amati et al (2006). The blue oval represents the thermal
component of GRB 060218 as recorded in the 0.3-10 KeV energy
channels of Swift (Campana et al.  2006). The lavender oval
represents the non-thermal component in the same energy range, which
we take to be the total minus the thermal component, both as
reported in Campana (2006). The quantity $E_{iso}$ [x axis] is
plotted in units of $10^{52}$ergs while $E_{peak}$ [y axis]
typically ranges from about 1 KeV to 1 MeV.
}{\label{f02}\footnotesize}
\end{figure}


I thank D. Guetta, C. Dermer, B. Katz, and E. Waxman for extensive conversations. I thank E. Golbraikh for technical assistance. I acknowledge support from the Israel-U.S. Binational Science Foundation, the Israel Science Foundation, and the Joan and Robert Arnow Chair of Theoretical Astrophysics.
\bigskip 

\noindent Amati, L. 2006,  M.N.R.A.S. 372, 233

\noindent Amati, L, Della Valle, M.  Frontera, F., Masulani, D.,  Guidorzi, C.,
Montenari, E., \&  Pian, E.  2007, A. \&  A., 463, 903

\noindent Band, D. 2002, ApJ 578, 806

\noindent Campana, S. et al., 2006, Nature, 442, 1008

\noindent Eichler, D. 1994, ApJS, 90, 877

\noindent Eichler, D., \& Levinson, A. 2004, ApJ, 614, L13

\noindent Eichler, D., Guetta, D. \& Pohl, M., 2010, ApJ, 722, 543

\noindent Guetta,D. \& Eichler, D., 2010, ApJ 710, 392

\noindent Levinson, A. 2006,  ApJ 648, 510

\noindent Levinson A. \& Eichler, 1993, ApJ 418, 386

\noindent Mandal, S. \& Eichler, D. 2010, ApJ, 713, L55

\noindent Nakar E. \& Piran, T. 2005, MNRAS, 360, L73

\noindent Paciesas, W.S. et al 1999, ApJS, 122, 465

\end{document}